\documentclass[10pt,twocolumn,showpacs]{revtex4}
\usepackage{amsmath}
\usepackage{bbold}
\usepackage{graphicx}

\def\Box{\kern1pt\vbox{\hrule height 1.2pt\hbox{\vrule width 1.2pt\hskip 3pt 
\vbox{\vskip 6pt}\hskip 3pt\vrule width 0.6pt}\hrule height 0.6pt}\kern1pt}

\def\be{\begin{equation}}
\def\ee{\end{equation}}
\def\bea{\begin{eqnarray}}
\def\eea{\end{eqnarray}}

\begin{document}


\title{Real space
tomography of the primordial Universe with cluster polarization}

\author{L. Raul Abramo}
\email{abramo@fma.if.usp.br}
\affiliation{Instituto de F\'{\i}sica,
Universidade de S\~ao Paulo \\
CP 66318, 05315-970, S\~ao Paulo, Brazil}

\author{Henrique S. Xavier}
\affiliation{Instituto de F\'{\i}sica,
Universidade de S\~ao Paulo \\
CP 66318, 05315-970, S\~ao Paulo, Brazil}

\begin{abstract}

We describe how a survey of the polarization of the cosmic microwave
background induced by Compton scattering in galaxy clusters can be 
used to make a full spatial reconstruction of
the primordial ($z \sim 1089$) matter distribution inside our 
surface of last
scattering. This ``polarization tomography" can yield a
spatial map of the initial state of the Universe just as
gravitational collapse was beginning to drive structure formation. 
We present a transparent method and simple formulas from which one 
can compute the 3D primordial map in real and in Fourier space, 
given a 3D map of the polarization due to galaxy
clusters. The advantage of the real space reconstruction is that
it is free from the statistical uncertainties which are inherent
in the Fourier space reconstruction.
We discuss how noise, partial sky covering and depth of the survey
can affect the results.



\end{abstract}

\pacs{98.65.Cw, 98.70.Vc, 98.80.-k, 98.80.Es}

\maketitle

\noindent
\underline{\bf Introduction}
The latest observations of the cosmic microwave background (CMB) \cite{WMAP1y,WMAP3y}
and of the galaxy distribution in the low-redshift Universe 
\cite{2dF,SDSS} have shown a spectacular agreement with the standard 
theory of structure formation \cite{Mukhanov,Dodelson}.
In this scenario, the initial conditions were set up by inflation,
which created a nearly scale-invariant spectrum of Gaussian density perturbations
with tiny variance ($\sigma \sim 10^{-5}$), and
all the visible 
structure we see today developed through gravitational instability from the
inflationary seeds.

However, these two sets of observations are made at two very 
different moments in time, and they probe very different regions of
space: while the CMB is formed essentially 
through the Sachs-Wolfe effect at the time of recombination ($z \sim 1089$),
galaxies on the other hand are observed only in relatively 
recent times ($z\lesssim 5$.)
Moreover, the CMB gives us a picture of the thin spherical
shell known as the last scattering surface (LSS),
which corresponds to the original location, at the time of recombination, 
of the CMB photons we observe on Earth today,
while the galaxy distribution, in practice, can only
be observed well inside our LSS.
This fact leads to a statistical ``leap of faith" in the argument for 
the standard 
scenario: we believe that the matter distribution {\it inside} our LSS is 
similar to the matter distribution {\it on} our LSS,
and therefore a typical configuration for the matter 
distribution at $z\sim 1089$ would lead to a map of the large-scale 
structure at $z \lesssim 5$
which has similar statistical properties to the one we actually observe.
This leap can be a particularly long one if the CMB 
anisotropies, which provide a picture of the 
matter distribution on our LSS, shows suspicious anomalies
\cite{OTZH,Efstathiou,Wagg,Copi04,Schwarz04,Hansen04,Eriksen04a,LM05,Armando05,Armando06,OT06,Copi06,Abramo06}. 

Kamionkowski and Loeb \cite{KL97} were the first to realize that it is 
possible, in principle, to recover information about the primordial 
matter distribution inside our LSS using cluster polarization of the 
CMB -- see also \cite{SS00,BCH,Portsmouth04,SP05,Bunn06}. 
This is because the CMB photons, as they Compton-scatter on free electrons in 
a galaxy cluster (the Sunyaev-Zel'dovich effect -- see \cite{SZ,SZrev}),
acquire a polarization which
reflects the quadrupole of the temperature distribution of the incident 
photons \cite{SS99}.
Hence, observing the polarization pattern in a galaxy cluster is tantamount to
observing the quadrupole of the CMB in the LSS seen by that cluster. 
In principle, this means that by sampling all the LSS's of clusters inside 
our LSS through cmb polarization, we could
reconstruct the temperature distribution of photons in the whole volume inside
our LSS, and thus reconstruct the matter distribution in that volume -- see Fig. 1. 

Notice
that a similar argument could be made for the CMB polarization which is 
induced at the time of last scattering through the local temperature 
quadrupole as well as for the polarization from the epoch of reionization. 
However, the former is
highly degenerate with the information we already have from the CMB 
temperature itself (see \cite{YW05} for a reconstruction scheme based on
the CMB temperature), and the latter is a large-scale effect which is
widely spread over the line of sight (as the ionization depth in this case is
only significant over horizon-scale distances), making it difficult to 
trace the polarized photons back to a particular spacetime event over
the past light-cone. Nevertheless, if reionization is patchy and inhomogeneous,
polarization combined with 21-cm observations can be
used much in the same way as cluster polarization to reconstruct the
primordial matter distribution \cite{HIM06} -- and formulas very similar 
to the ones we show here can be used in that case as well.

In practice, the reconstruction scheme based on cluster polarization
has been difficult to implement for several
reasons. First and foremost, because the amplitude of the polarized
signal is at most a few percent of the temperature fluctuation signal
for typical cluster optical depths \cite{SS99,LSA04,AW04,CBS04}.
Second, foregrounds such as lensing \cite{ZS98}, 
extended galactic emissions \cite{Page06,Reich,Leonardi}, 
cluster peculiar velocities \cite{CFL00,SRON06} and 
filamentary structures \cite{LSA04}, can be hard to extricate from the data. 
And third, the methods for recovering the matter distribution
from polarization maps lacked transparency 
\cite{SS00,Portsmouth04,SP05}.

We believe that these difficulties will be overcome.
First, the latest technological advances have made it 
possible to start searching for the B-mode of polarization 
induced by gravitational waves from inflation 
\cite{EBEX,Polarbear,Yoon06,Spider} -- which will be more than enough 
sensitivity to detect polarization at the levels found in clusters.
Optic and X-ray data, which are crucial to
determine the optical depths and redshifts of the clusters, should make 
de-lensing significantly easier at the arcminute scales we are 
interested \cite{KCK02,SH03}.
Filamentary structures can be avoided by cutting out regions of the sky with 
very high gas columns or projected surface density of clusters. Similarly,
galactic foregrounds can be either directly subtracted with multi-frequency
observations, or a mask can be applied onto the maps. 
Cluster peculiar velocities can also be dealt with in multi-frequency
surveys \cite{BCH}. 

As for the reconstruction of the matter distribution, we will present
a transparent method and simple formulas for the reconstruction of 
the matter distribution both in Fourier as well as in real space. We
have considered the contributions from the Sachs-Wolfe and the
integrated Sachs-Wolfe effects (SWe and ISWe, respectively). It is trivial
to consider surveys with partial sky covering, as well as any type
of redshift binning one may employ. We also study the effect of 
adding noise to the polarization data and its impact on the reconstructed
maps.

\vskip 0.3cm

\noindent
\underline{\bf Cluster polarization}
Since only the Stokes parameters $Q$ and $U$ are relevant in
cosmology \cite{ZS97,KKS97,HW97}, we define the dimensionless 
complex polarization of a cluster at redshift $z$ in the
direction $\hat{n}$ as:
\be
\label{Polar}
P(z,\hat{n}) \equiv \frac{Q(z,\hat{n}) - i \, U(z,\hat{n})}{T(z) \tau_{z,\hat{n}}} \; ,
\ee
where $T(z)= 2.726 (1+z)$ K and $\tau_{z,\hat{n}}$ is the optical depth 
for the cluster at redshift $z$ and position $\hat{n}$.

At linear order in the cosmological perturbations (we neglect the 
Riess-Sciama effect), the polarization for a given cluster depends on
the gravitational potential $\Phi$ {\it on} the cluster's LSS through the 
SWe, and it depends on the fluctuations of the potential {\it inside} the 
cluster's LSS through the ISWe -- see Fig. 1.

\begin{figure}
\includegraphics[width=4.5cm]{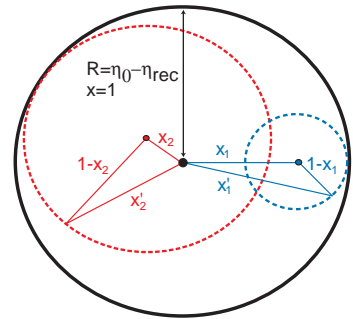}
\caption{\label{fig:1}
Last scattering surfaces seen by an observer at the origin at $\eta = \eta_0$ 
(solid black circle) and by two clusters at times $\eta_1$ and $\eta_2$ 
(dashed grey circles -- dashed red and blue circles in the online version.)
The variable $x$ is physical distance in units of the distance to our
LSS, $x=(\eta-\eta_{rec})/(\eta_0-\eta_{rec})$. The distance from the
origin to the LSS of a given cluster at $x$
spans the interval $|1-2x| \leq x' \leq 1$. 
}
\end{figure}

Usually one describes the gravitational potential in terms of its
Fourier spectrum:
\be
\label{Phi_k}
\Phi [\vec{r}(z,\hat{n})] = \int \frac{d^3 k}{(2 \pi)^{3/2}}
e^{-i \vec{k} \cdot \vec{r}} \tilde\Phi (\vec{k}) \; .
\ee
The Fourier transform of the gravitational potential has Gaussian
random P.D.F.'s set up by inflation, with
$\langle \tilde{\Phi}(\vec{k}) \tilde{\Phi}^*(\vec{k}') \rangle
= k^{-3} P(k) \delta (\vec{k}-\vec{k}')$, and $P(k) = A k^{n_s-1}$
a nearly scale-invariant function.

For this problem it is very useful to describe the gravitational potential 
in the volume inside our LSS (which includes the LSS's of all clusters) in 
the following way:
\be
\label{Phi_x}
\Phi (\vec{r}) = \sum_{\ell,m} b_{\ell m} (r) \,
Y_{\ell m} (\hat{n}) \; ,
\ee
where $r = \eta_0-\eta (z) $ and $\eta$ is conformal time. 
The coefficients $b_{\ell m} (r)$ are
well-behaved functions of the radius if $\Phi(\vec{r})$ is a smooth
function.
Since the radius of our LSS, at $z_{\rm rec} \simeq 1089$, is
$R_{\rm LSS} =\eta_0-\eta_{\rm rec}$, it is useful to write all physical 
distances in terms of this maximal radius, $x \equiv r/R_{\rm LSS}$ so
$x$ lies in the interval $0 \leq x \leq 1$.

We can write $\tilde\Phi (\vec{k})$ in the same way
as Eq. (\ref{Phi_x}), in terms of coefficients 
$\tilde{b}_{\ell m} (k)$.
The formal relationship between $b_{\ell m} (x)$ and 
$\tilde{b}_{\ell m} (k)$ is a Hankel transform:
\be
\label{b_btilde}
\tilde{b}_{\ell m} (k) = \sqrt{\frac{2}{\pi}} \, (-i)^\ell \int dx \, 
x^2 \, j_\ell (k x) \,
b_{\ell m} (x) \; .
\ee
The inverse relation is found by exchanging $k \leftrightarrow x$ and
multiplying by $(-1)^\ell$.
The fact that $\Phi (\vec{r})$ is real translates into the conditions
$\tilde{b}^*_{\ell m} = (-1)^{\ell+m} \tilde{b}_{\ell, -m}$
and ${b}^*_{\ell m} = (-1)^{m} {b}_{\ell, -m}$.

A decomposition analogous to Eq. (\ref{Phi_x}) can be made for the 
cluster polarization as well, except that here the spin-2 character of the
polarization field demands that we use the spin-weighted 
spherical harmonics instead \cite{Bunn06}:
\be
\label{p_Y}
P = \sum_{\ell,m} p_{\ell m} (x) \,
{}_{2} Y_{\ell m} (\hat{n}) \; .
\ee

In terms of the coefficients $p_{\ell m} (x) $ and
$\tilde{b}_{\ell m} (k)$, the cluster
polarization takes on a very simple expression:
\be
\label{p_bk}
p_{\ell m}(x) = 
\int dk \, \tilde{K}_\ell (x,k) \, \tilde{b}_{\ell m} (k) \; ,
\ee
where the Fourier space Kernel $\tilde{K}_\ell(x,k)$ is given by:
\be
\label{Kernel_k}
\tilde{K}_\ell (x,k) = i^\ell \, 60 \pi \sqrt{\frac32} \, 
k^{2} \, f_\ell(kx) \, \tilde{\Delta}_2 (x,k) \; ,
\ee
with  \cite{Bunn06}:
\bea
\label{fl}
f_\ell (kx) &=& - \frac{1}{45} \sum_{\lambda=\ell-2,\ell,\ell+2} 
(-1)^{(\lambda-\ell)/2} (2\lambda +1) \\ \nonumber
&\times&
\left( 
\begin{array}{ccc}
2 & \ell & \lambda \\
2 & -2 & 0
\end{array}
\right)
\left( 
\begin{array}{ccc}
2 & \ell & \lambda \\
0 & 0 & 0
\end{array}
\right)
j_\lambda(kx) \; ,
\eea
and
\be
\label{Delta2}
\tilde{\Delta}_2 (x,k) = 
j_2 [k(1-x)] - 6 \int_x^1 dx'' j_2 [k(x''-x)] \frac{d G(x'')}{dx''}  \; .
\ee
The first and second terms in
Eq. (\ref{Delta2}) stand for, respectively, the SWe
and the ISWe. The growth function
$G(x)=D[z(x)]/a[z(x)]$ is normalized to unity today and its derivative
is non-negligible only in relatively recent times ($z\lesssim 5$)
as dark energy becomes more relevant.

In fact, it turns out that the sum in Eq. (\ref{fl}) is 
a particular realization of the spherical Bessel function's 
recursion relations, and the result is:
\be
\label{fl2}
f_\ell (kx) = \sqrt{\frac{(\ell+2)!}{6 (\ell-2)!}} \, \frac{j_\ell(kx)}{30 (kx)^2} \; .
\ee
This window function, which mediates the exchange of power between 
the ``orbit" angular momenta $\ell\geq 2$ and the spin angular momentum 
of the polarization field,
is identical to the one found in Ref. \cite{SP05} for the 
correlation function of the polarization coefficients, and 
it ensures that the power is 
conserved, $\sum_{\ell=2}^{\infty} (2\ell+1) f_\ell^2=1$.
It is also the same (up to a normalization factor) as the radial function
$\epsilon^{(0)}_\ell$ of \cite{HW97}, which shows that
the cluster-induced polarization is a pure $E$ (or gradient) mode.

\vskip 0.3cm

\noindent
\underline{ \bf Polarization tomography in real space}
Even perfect knowledge of the polarization field on the
whole volume of our LSS does not eliminate the statistical
errors in the Fourier space coefficients $\tilde{b}_{\ell m} (k)$
-- as Eq. (\ref{b_btilde}) is cut-off at $x=1$.
These uncertainties 
are a 3D analog of the 2D ``cosmic variance" 
that plagues CMB anisotropies. However, even though these statistical 
uncertainties can be large, especially for the longest-wavelength 
modes, we can still reconstruct the 3D spatial map of the
primordial fluctuations inside the
LSS with as much accuracy as the observations allow.
This is a nontrivial statement, as the polarization we
observe from a cluster carries only the information of 
the temperature quadrupole as seen by that cluster,
projected along the line of sight. Nevertheless, it turns out that 
this is just enough information to allow (at least in theory) for a complete
spatial reconstruction.

Plugging Eq. (\ref{b_btilde}) into Eq. (\ref{p_bk}) and using 
a generalization of the Weber-Shafheitlin integral \cite{Watson}
we obtain:
\be
\label{p_x_Kernel}
p_{\ell m} (x) = \int_0^1 \, dx' \, K_\ell(x,x') b_{\ell m} (x') \; ,
\ee
where the real space kernel $K_\ell(x,x')$ is given by an SWe piece 
and an ISWe piece. It turns out that $K_\ell$ is real, which is
another manifestation of the fact that cluster polarization is a pure 
$E$ mode.

The SWe kernel is:
\bea
\nonumber
K^{(SW)}_\ell &=& \frac{\pi^2}{2} \sqrt{\frac{(\ell+2)!}{2 \pi (\ell-2)!}} \, 
\frac{{x'}^3}{x(1-x)^3} 
\\ 
\nonumber
&\times&
\theta ( |x'+x| - |1-x|)] \,
\theta ( |1-x| - |x'- x|)
\\ 
\label{K_SW}
&\times&
\sin^2 \psi \, P_\ell^{\, -2} (\cos\psi)
\; ,
\eea
where $P_\ell^{\, n}$ is the generalized Legendre polynomial,
and $\cos\psi = [x^2+x'^2-(1-x)^2]/(2x x')$ is the cosine 
of the angle between $x$ and $x'$ in a triangle with 
sides $x$, $x'$ and $1-x$ -- see Fig. 1. The
step-functions in Eq. (\ref{K_SW}), which 
vanish unless $|1-2x| \leq x' \leq 1$,
automatically imply causality and
ensure that the polarization in each point
can only depend, through the SWe, on
the gravitational potential on the LSS of that point.

The ISWe part is given by:
\bea
\nonumber
K^{(ISW)}_\ell &=& - 3 \pi^2  
\sqrt{\frac{(\ell+2)!}{2 \pi (\ell-2)!}} \, \frac{{x'}^3}{x}
\int_x^1 dx'' \frac{1}{(x''-x)^3} 
\\ \nonumber
&\times&
\theta ( |x'+x| - |x''-x|)] \,
\theta ( |x''-x| - |x'- x|)
\\ 
\label{K_ISW}
&\times& \frac{dG(x'')}{dx''} \,
\sin^2 \psi' \, P_\ell^{\, -2} (\cos\psi')
\; ,
\eea
where $\cos\psi' = [x^2+x'^2-(x''-x)^2]/(2x x')$ is the cosine of the
angle between $x$ and $x'$ in the triangle with sides $x$, $x'$ and $x''-x$.
As was the case for the SWe kernel above, 
it is straightforward to show from the step-functions in Eq. (\ref{K_ISW}) 
that the ISWe kernel vanishes if $x' > 1$.
Detailed geometric interpretations of the 
real space kernels can be found in a forthcoming publication \cite{Nois07}.

Notice that by virtue of the step-functions in the kernels above,
the variables $x$, $x'$ and $x''$ all lie in the interval $[0,1]$,
so both $p_{\ell m}$ and $b_{\ell m}$ have compact support.
This means that the linear problem of finding the coefficients 
$b_{\ell m} (x')$ given the data $p_{\ell m} (x)$ is completely 
well-defined in real space. This is in contrast with the problem 
of finding the Fourier space coefficients $\tilde{b}_{\ell m}$
given the same data: because the $p_{\ell m}$'s in Eq. (\ref{p_bk})
can only be integrated over the interval $0\leq x \leq 1$, the inversion
of Eq. ({\ref{p_bk}) is singular. Evidently, this is a manifestation of 
the statistical uncertainties of the $\tilde{b}_{\ell m}$'s.

Eqs. (\ref{p_x_Kernel})-(\ref{K_ISW}) are 
the main results of this paper. Next we discuss applications of these
formulas.

\begin{figure}
\includegraphics[height=7cm]{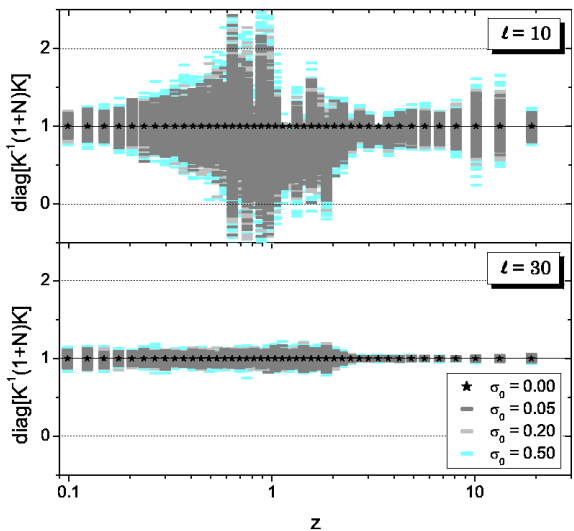}
\caption{\label{fig:2}
Noisy estimator $K_\ell^+ (\mathbb{1}+N_\ell) K_\ell$ 
for $\ell=10$ (top) and $\ell=30$ (bottom), assuming zero
errors (black stars) and assuming Gaussian errors
with variances $\sigma_0=0.05,0.2,0.5$ (boxes of dark, medium
and light grey.) 
We used 500 realizations of $N_\ell$ for each $\sigma_0$, and
we use $f_{\rm sky}=2/3$.
In this example we have assumed a hypothetical survey with 100 
redshift bins up to $z=10$, and we recover information on 50 
redshift bins.
}
\end{figure}

\vskip 0.3cm

\noindent
\underline{\bf Map reconstruction}
In practice, the integrals (\ref{p_bk}) and (\ref{p_x_Kernel}) become 
sums over bins, so we have to solve a linear problem of the type 
$P_{\ell m \,  (i) } = \sum_j K_{\ell \,  (ij)} B_{\ell m \,  (j)}$,
with $K_{\ell}$ some matrix that need not be square if we want to
compute less parameters $B_{\ell m \, (j)}$ than there are data points 
$P_{\ell m \, (i)}$.
But even if the matrix $K_\ell$ is square, in general it is
singular. This is a severe problem for the Fourier space
kernel, which has many singular eigenvalues 
due to
the statistical uncertainties in the long-wavelength modes
$\lambda \gtrsim R_{\rm LSS}$ \cite{Nois07}. For the real space kernel
we find that there are essentially no singular eigenvalues, which
allows, at least in principle, for an exact inversion.

For linear problems which involve rectangular or singular square
matrices, we can employ the pseudo-inverse \cite{Penrose}.
The pseudo-inverse $K^+$ of a matrix $K$ 
reduces to the inverse in the case of square
non-singular matrices, and it gives the least-squares 
solution $\hat{B}=K^+ \cdot P$ to the linear problem 
$P = K \cdot B$.
In this framework it is trivial to consider noisy data: the 
observations become $P_{\ell m} \rightarrow P^N_{\ell m} 
= ({\mathbb{1}}+N_\ell) \cdot P_{\ell m}$, 
where $N_\ell$ is the noise matrix which is 
diagonal if the errors are uncorrelated.

In the presence of noisy data the estimator $\hat{B}_{\ell m}$ 
becomes a noisy estimator $\hat{B}^N_{\ell m}$, which can be
related to the actual parameters $B_{\ell m}$ that we want to
measure by:
\be
\label{B_N}
\hat{B}^N_{\ell m} = K_\ell^+ (\mathbb{1}+N_\ell) K_\ell \cdot B_{\ell m} \; .
\ee
Here $N_\ell$ is a Gaussian random variable with zero mean and variance given by:
\be
\label{sigma}
\sigma_\ell \approx \sqrt{\frac{2}{2\ell+1}  \frac{1}{f_{\rm sky}} }
\left( 1 - f_{\rm sky} + \sigma_0^2 \right)^{1/2} \; ,
\ee
where $\sigma_0$ includes factors such as instrument noise, number of pixels and
redshift, 
and $f_{\rm sky} \leq 1$ is the fraction of the sky mapped by that instrument.
The term $f_{\rm sky}$ inside the parenthesis in Eq. (\ref{sigma}) subtracts
cosmic variance, which is not relevant for 
the accuracy of the actual 
map reconstruction.

Since for the real space kernel $K^+ K$ is nearly identical to the 
unit matrix (unlike the case of the Fourier space kernel), one can 
easily find that the covariance matrix of
the estimator $\hat{B}^N_{\ell m}$ is given by:
\be
\label{covariance}
Cov(\hat{B}^N_{\ell m})_{i j} \simeq \sigma_\ell^2 \delta_{i j} 
\sum_k B_{\ell m \, k} B^*_{\ell m \, k} \; .
\ee
One can recognize the sum over redshift bins in the r.h.s. of Eq. (\ref{covariance}) as the
projected $C_\ell$ -- i.e., $\int dz \, C_\ell(z)$.

If the matrix $K_\ell^+ (\mathbb{1}+N_\ell) K_\ell$ 
is close to a unit matrix, then the noisy estimator will be a
good approximation to the true parameters. We can evaluate 
the goodness of the noisy estimator by looking at the 
diagonal of that matrix. 

We have done this for all the modes between
$\ell=2$ and $\ell=50$, using errors 
$\sigma_0=0.05,0.2,0.5$, in
a hypothetical survey up to $z=10$ with 100 bins, and in which
we reconstruct the $b_{\ell m}$'s in 50 redshift bins. 
In this example, for $\ell=2$ the reconstruction
fails, mainly because of poor sky coverage. However, as $\ell$ increases 
sky coverage becomes less of a problem and the reconstruction becomes
progressively better. Fig. 2 shows the accuracy of the reconstruction
for $\ell=10$ and $\ell=30$.
A deeper survey does not significantly improve the reconstruction,
but a survey shallower than $z \lesssim 3$ impacts negatively the
reconstruction at low redshifts, as the LSS's of low-redshift clusters
are entirely at high redshifts. The severest 
limitation to the reconstruction scheme is sky coverage.

\vskip 0.3cm

\noindent
\underline{ \bf Conclusions}
We have shown how to reconstruct the 
primordial density field inside our LSS
using cluster polarization.
In principle, a complete reconstruction is possible with
this method. In practice, sky coverage and noise limit the
accuracy of the reconstruction, especially for low $\ell$'s and
for intermediate redshifts ($z \sim 1$). Nevertheless,
a reconstruction of the primordial density field in the
local Universe ($z \lesssim 0.5$) and on angular scales $2^{\rm o} \lesssim 
\theta \lesssim 20^{\rm o}$, is feasible with the next
generation of polarimeters which are being used to search for
the $B$ mode from gravitational waves.

The primordial field is highly correlated with
the present matter distribution. 
Direct comparison of the two is a key test of
our models of structure formation. 
In particular, correlating the primordial map with the present 
locations of clusters and superclusters tests the
evolution of structures, as potential wells only move due to 
nonlinear effects. Finally, cross-correlating the
primordial map with tracers of the growth of structure 
is a new test that constrains the parameters of dark energy.

\vskip 0.3cm

\noindent
\underline{Acknowledgments:} 
R.A. would like to thank N. Aghanim, 
J. C. A. Barata, M. Coutinho-Neto, T. Villela and C. A. Wuensche
for valuable discussions. 
Financial support for this work has been provided by
FAPESP and CNPq.


\begin{thebibliography}{99}

\bibitem{WMAP1y} C. L. Bennett {\it et al.}, Astrophys. J. Suppl.
Ser. {\bf 148}, 1 (2003), astro-ph/0302207;
D. Spergel {\it et al.}, Astrophys. J. Suppl. 
Ser. {\bf 148}, 175 (2003), astro-ph/0302209. 

\bibitem{WMAP3y} D. Spergel {\it et al.}, arXiv: astro-ph/0603449.

\bibitem{2dF} W. Percival {\it et al.},
{\it Mon. Not. Roy. Astron. Soc.} {\bf 327}: 1297 (2001),
astro-ph/0105252;
J. Peacock {\it et al.}, {\it Nature} {\bf 410}: 169 (2001),
astro-ph/0103143;
Licia Verde {\it et al.}, {\it Mon. Not. Roy. Astron. Soc.}
{\bf 335}: 432 (2002), astro-ph/0112161.

\bibitem{SDSS} 
K. Abazajian {\it et al.}, {\it Astron. J.} {\bf 126}: 2081 (2003),
astro-ph/0305492;
M. Tegmark {\it et al.}
{\it Phys. Rev.} {\bf D69}: 103501 (2004), 
astro-ph/0310723;
U. Seljak {\it et al.}, 
{\it Phys. Rev.} {\bf D71}: 103515 (2005),
astro-ph/0407372.


\bibitem{Mukhanov} V. Mukhanov, ``Physical Foundations
of Cosmology'' (Cambridge University Press, 2005).

\bibitem{Dodelson} S. Dodelson, ``Modern Cosmology''
(Academic Press, 2004).



\bibitem{OTZH} A. de Oliveira-Costa, M. Tegmark, M. Zaldarriaga and A. 
Hamilton, {\it Phys. Rev.} {\bf D69}: 063516 (2004), astro-ph/0307282.

\bibitem{Efstathiou} G. Efstathiou, 
{\it Mon. Not. Roy. Astron. Soc.} {\bf 348}: 885 (2004), astro-ph/0310207.

\bibitem{Wagg} E. Gazta\~naga {\it et al.},
{\it MNRAS} {\bf 346}: 47 (2003), astro-ph/0304178.

\bibitem{Copi04} C. J. Copi, D. Huterer, and G. D. Starkman,
{\it Phys. Rev.} {\bf D70}, 043515 (2004), astro-ph/0310511.

\bibitem{Schwarz04} D. Schwarz, G. Starkman, D. Huterer and C. Copi,
{\it Phys. Rev. Lett.} {\bf 93}: 221301 (2004), astro-ph/0403353; 
C. Copi, D. Huterer, D. Schwarz and G. Starkman, 
{\it Mon. Not. Roy. Astron. Soc.} {\bf 367}:79 (2006), astro-ph/0508047.

\bibitem{Hansen04} F. Hansen, P. Cabella, D. Marinucci and N. Vittorio,
{\it Astrophys. J.} {\bf 607}: L67 (2004), astro-ph/0402396.

\bibitem{Eriksen04a} H. Eriksen, F. Hansen, A. Banday, K. G\'orski and
P. Lilje, {\it Astrophys. J.} {\bf 605}: 14 (2004), astro-ph/0307507.

\bibitem{LM05} K. Land and J. Magueijo, 
{\it Phys. Rev. Lett.} {\bf 95}: 071301 (2005 ), astro-ph/0502237;
{\it Mon. Not. Roy. Astron. Soc.} {\bf 362}: 838 (2005), astro-ph/0502574.

\bibitem{Armando05} A. Bernui, B. Mota, M. Rebou\c{c}as and R. Tavakol,
astro-ph/0511666.

\bibitem{Armando06} A. Bernui, T. Villela, C. Wuensche, R. Leonardi and I. Ferreira,
astro-ph/0601593.

\bibitem{OT06} A. de Oliveira-Costa and M. Tegmark, arXiv: astro-ph/0603369.

\bibitem{Copi06} C. Copi, D. Huterer, D. Schwarz, and G. 
Starkman, arXiv: astro-ph/0605135.

\bibitem{Abramo06} L. R. Abramo {\it et al.}, 
{\it Phys. Rev.} {\bf D74}: 063506 (2006), astro-ph/0604346.


\bibitem{KL97} M. Kamionkowski and A. Loeb, 
{\it Phys. Rev.} {\bf D56}: 4511 (1997), astro-ph/9703118.

\bibitem{SS00} N. Seto and M. Sasaki, {\it Phys. Rev.} {\bf D62}: 123004
(2000), astro-ph/0009222.

\bibitem{BCH} A. Cooray and D. Baumann,
{\it Phys. Rev.} {\bf D67}: 063505 (2003), astro-ph/0211095;
A. Cooray, D. Huterer and D. Baumann,
{\it Phys. Rev.} {\bf D69}: 027301 (2004), astro-ph/0304268;
D. Baumann and A. Cooray, 
{\it New Astron. Rev.} {\bf 47}: 839 (2003), astro-ph/0304416.


\bibitem{Portsmouth04} J. Portsmouth, {\it Phys. Rev.} {\bf D70}: 063504
(2004), astro-ph/0402173.

\bibitem{SP05} N. Seto and E. Pierpaoli, {\it Phys. Rev. Lett.} {\bf 95}: 
101302 (2005), astro-ph/0502564.


\bibitem{Bunn06} E. Bunn, {\it Phys. Rev.} {\bf D73}: 123517 (2006)
astro-ph/0603271.



\bibitem{SZ} R. Sunyaev and Ya. B. Zel'dovich,
{\it Astrophys. Space Sci.} {\bf 7}: 3 (1970); 
{\it Annu. Rev. Astron. Astrophys.} {\bf 18}: 537 (1980).

\bibitem{SZrev} 
Y. Rephaeli, {\it Annu. Rev. Astron. Astrophys.} {\bf 33}: 541 (1995); 
M. Birkinshaw, {\it Phys. Rep.} {\bf 310}: 97 (1999);
J. Carlstrom, G. Holder and E. Reese, {\it Annu. Rev. Astron. Astrophys.}
{\bf 40}: 643 (2002).

\bibitem{SS99} S. Sazonov and R. Sunyaev, {\it MNRAS} {\bf 310}: 765 (1999).

\bibitem{YW05} A. Yadav and B. Wandelt, {\it Phys. Rev.} {\bf D71}: 123004
(2005), astro-ph/0505386.

\bibitem{HIM06} G. Holder, I. Iliev and G. Mellema, astro-ph/0609689.


\bibitem{LSA04} G.-C. Liu, A. da Silva and N. Aghanim, {\it Astrophys. J.} 
{\bf 621}: 15 (2005),
astro-ph/0409295.

\bibitem{AW04} A. Amblard and M. White, {\it New Astron.} {\bf 10}: 417 (2005), 
astro-ph/0409063.

\bibitem{CBS04} A. Cooray, D. Baumann and K. Sigurdson, astro-ph/0410006.


\bibitem{ZS98} M. Zaldarriaga and U. Seljak, {\it Phys. Rev.} {\bf D58}:
023003 (1998), astro-ph/9803150.

\bibitem{Page06}
L. Page {\it et al.}, astro-ph/0603450.

\bibitem{Reich} W. Reich, astro-ph/0603465.

\bibitem{Leonardi} R. Leonardi {\it et al.}, {\it New Astron. Rev.}
{\bf 50}: 977 (2006).


\bibitem{CFL00} A. Challinor, M. Ford and A. Lasenby,
{\it MNRAS} {\bf 312}: 159 (2000), astro-ph/9905227.

\bibitem{SRON06} M. Shimon, Y. Rephaeli, B. O'Shea and M. Norman,
{\it Mon. Not. Roy. Astron. Soc. } {\bf 368}: 511 (2006),
astro-ph/0602528.


\bibitem{EBEX} P. Oxley {\it et al.}, {\it Proceedings of the SPIE} 
{\bf 5543}: 320 (2004), astro-ph/0501111.

\bibitem{Polarbear} M. Myers {\it et al.}, {\it Appl. Phys. Lett.}
{\bf 86}: 114103 (2005), http://bolo.berkeley.edu/polarbear/ .

\bibitem{Yoon06}
K. Yoon {\it et al.}, astro-ph/0606278; 
B. Keating {\it et al.}, {\it Proceedings of the SPIE} {\bf 4843} 
(2003); http://www.astro.caltech.edu/~lgg/bicep\_front.htm.

\bibitem{Spider} http://www.astro.caltech.edu/~lgg/spider\_front.htm .


\bibitem{KCK02} M. Kesden, A. Cooray and M. Kamionkowski,
{\it Phys. Rev. Lett.} {\bf 89}: 011304 (2002), astro-ph/0202434.

\bibitem{SH03}
C. Hirata and U. Seljak, {\it Phys. Rev.} {\bf D68}: 083002 (2003);
astro-ph/0306354;
U. Seljak, C. Hirata, {\it Phys. Rev.} {\bf D69} 043005 (2004),
astro-ph/0310163.


\bibitem{ZS97} M. Zaldarriaga and U. Seljak, {\it Phys. Rev.} 
{\bf D55}: 1830 (1997), astro-ph/9609170.

\bibitem{KKS97} M. Kamionkowski, A. Kosowsky and A. Stebbins, {\it Phys. Rev.} 
{\bf D55}: 7368 (1997), astro-ph/9611125.

\bibitem{HW97} W. Hu and M. White,
{\it Phys. Rev. } {\bf D56}: 596 (1997), astro-ph/9702170.

\bibitem{Watson} G. Watson, ``A Treatise on the Theory of Bessel Functions"
(Cambridge U. Press, 1966).

\bibitem{Nois07} L. R. Abramo and H. Xavier, to appear.

\bibitem{Penrose} R. Penrose, {\it Proc. Cambridge Phil. Soc.} {\bf 51}: 406 (1955).

\end{thebibliography}
\end{document}